\documentclass[aps,pre,showpacs,twocolumn,groupedaddress,superscriptaddress]{revtex4}
\usepackage{graphicx}%
\usepackage{amsmath}
\usepackage{amssymb}
\usepackage{dcolumn}% Align table columns on decimal

\begin{document}

\title{Chimera and globally clustered chimera: Impact of time delay}

\author{Jane H.~Sheeba}%
\affiliation{Centre for Nonlinear Dynamics, School of Physics,
Bharathidasan University, Tiruchirappalli - 620 024, Tamilnadu, India}

\author{V.~K.~Chandrasekar}%
\affiliation{Centre for Nonlinear Dynamics, School of Physics,
Bharathidasan University, Tiruchirappalli - 620 024, Tamilnadu, India}

\author{M.~Lakshmanan}%
\affiliation{Centre for Nonlinear Dynamics, School of Physics,
Bharathidasan University, Tiruchirappalli - 620 024, Tamilnadu, India}

%\date{\today}

\begin{abstract}
Following a short report of our preliminary results [Phys. Rev. E 79, 055203(R) (2009)], we present a more detailed study of the effects of coupling delay in diffusively coupled phase oscillator populations. We find that coupling delay induces chimera and globally clustered chimera (GCC) states in delay coupled populations. We show the existence of multi-clustered states that act as link between the chimera and the GCC states. A stable GCC state goes through a variety of GCC states, namely periodic, aperiodic, long-- and short--period breathers and becomes unstable GCC leading to global synchronization in the system, on increasing time delay. We provide numerical evidence and theoretical explanations for the above results and discuss possible applications of the observed phenomena.
\end{abstract}



\maketitle

\section{Introduction}

Kuramoto, Battogokh and Shima discovered \cite{Kuramoto:02,Kuramoto:03,Shima:04} an interesting spatiotemporal
pattern which was later named \emph {chimera} by Abrams and Strogatz \cite{Abrams:04,Abrams:06}. The name
\emph{chimera}, which literally refers to something that is composed of seemingly
incompatible or incongruous parts, was coined for this phenomenon because a group of
identical oscillators splits into two groups of completely different character. Since its
discovery \cite{Kuramoto:02,Kuramoto:03,Abrams:04}, various theoretical and
numerical developments have been reported on the stability of chimera
states and their existence in systems with varied structures \cite{Abrams:04,Abrams:08},
including time delay \cite{Sethia:08}. It is clear that the chimera state cannot be
attributed to partial synchronization. The occurrence of partial synchronization in a population of non-identical oscillators is not surprising. 
On the other hand, if an identical group of oscillators splits into synchronized and desynchronized groups, it is called chimera.
Therefore, the discovery of chimera came as a surprise in the study of synchronization phenomenon in complex systems.

By and large, synchronization in coupled oscillator systems has been analytically and numerically investigated in a rigorous
manner over the past years \cite{Winfree:67,Strogatz:01}.
Possible routes to global synchronization and methods to control
synchronization have also been proposed \cite{Sherman:92,Rosenblum:04}.
However, complete understanding of the effects induced by coupling delay
in synchronization of coupled oscillator systems is still an open problem.
It is well known that time delay occurs in real physical systems. For example, in a
network of neuronal populations, there is certainly a significant
delay in propagation of signals. In addition there can also be
synaptic and dendritic delays. Other examples include finite reaction times of chemicals and
finite transfer times associated with the basic mechanisms that regulate gene transcription and mRNA
translation. 

The nature of coupling in coupled oscillator systems has been conventionally considered as
instantaneous during earlier studies. One of the main reasons for this assumption
is that it substantially simplifies the analysis of the system. In addition, such an approximation is more often physically
justified. However, the fact is that the consideration of time delay is vital for modeling real life systems. Furthermore,
as we will demonstrate in this paper, certain interesting dynamical phenomena in complex systems are characteristic of time delay and
they will not occur in systems without time delay. Since the introduction of time delay increases the effective dimension of the system, one can
expect certain complex phenomena to be explained in a better way in models of real physical systems when delay is included.

\begin{figure}
\begin{center}
\includegraphics[width=8.5cm]{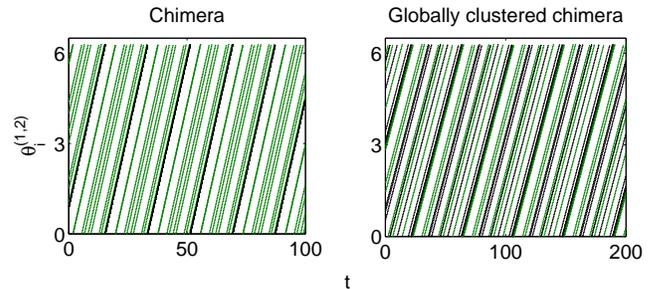}
\caption{(Color online) Occurrence of stable chimera and GCC states in system (\ref{chim_del01}).
Black and green (grey) lines represent oscillators in the first
and the second populations, respectively. Here $\{f,h\}=\{\sin(\theta),\cos(\theta)\}$, $\tau_1=n\tau_2=n\tau$ with $n=0$,
$A=0.4$, $B=0.6$ and $\tau=2$ for the chimera and $\tau=4$ for the GCC.}
\label{Motiv}
\end{center}
\end{figure}

In this paper, following our Rapid Communication \cite{Jane:09b}, we present a more detailed discussion of the effects of coupling delay in inducing chimera and globally clustered chimera (GCC) states in systems of coupled identical oscillator populations. By a GCC state, here we mean a state where the system splits into two different groups, one synchronized and the other desynchronized, each group comprised of oscillators
from both the populations. Since a global clustering (mixing) of oscillators from both the populations occur in this case, we call this
state a GCC. This is different from the chimera state where one of the populations is synchronized while the
other is desynchronized \cite{Abrams:04} (See Fig. \ref{Motiv} for an illustration. Fuller details are given in Sec. \ref{NumStud}). In addition, we find the existence of multi--clustered chimera and GCC states that are induced by time delay. In the multi--clustered states there are more than one synchronized groups (that contain oscillators from the same population in the case of chimera and from different populations in the case of GCC) and the rest of the oscillators in the populations are desynchronized. We present a detailed possible analytical explanation for the numerically observed phenomena.

The paper is organized as follows: In Sec. \ref{NumStud} we explain the numerical methods and analysis carried out. We also explain the numerical method to calculate the modified order parameter. In Sec. \ref{Breat}, we discuss the breathing and unstable nature of the chimera and GCC states and explain the different types of breathers that occur in the system under study. We discuss the existence of multi-clustered states in Sec. \ref{Multi}. Sec. \ref{Anal} provides analytical evidence and support of the numerical results discussed in the paper. We discuss possible applications of the chimera and GCC states in real physical systems in Sec. \ref{App}. Finally in Sec. \ref{Sum} we summarize the results of the paper.

\section{Numerical studies}
\label{NumStud}
Let us consider a system of two populations of identical oscillators coupled through a finite delay, represented
by the following equation of motion
\begin{eqnarray}
\dot{\theta_i}^{(1,2)}&=& \omega - \frac{A}{N}
\sum_{j=1}^{N}
f(\theta_i^{(1,2)}(t)-\theta_j^{(1,2)}(t-\tau_1)) \nonumber \\
&&-
\frac{B}{N}\sum_{j=1}^{N}
h(\theta_i^{(1,2)}(t)-\theta_j^{(2,1)}(t-\tau_2)), \nonumber \\
&&\qquad \qquad \qquad \qquad \qquad i=1, 2,\ldots, N.
\label{chim_del01}
\end{eqnarray}

\begin{figure}
\begin{center}
\includegraphics[width=4.5cm]{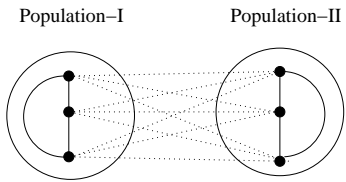}
\caption{Schematic representation of system (\ref{chim_del01}) with $N=3$ comprised of two populations of all--to--all coupled oscillators; the oscillators within each population are identical. Here solid lines represent coupling within a population (with strength $A$) and dotted lines represent coupling between populations (with strength $B$).}
\label{sys_schm}
\end{center}
\end{figure}
Here $\omega$ is the natural frequency of the oscillators in the populations and it is the same for
all the oscillators in both the populations making all of them identical. However, in order to differentiate one population from the other we set the initial distribution of the phases
of the first population to be uniformly distributed between $0$ and $\pi$  and that of the second population to be uniformly distributed between $\pi$ and
$2\pi$. The coupling strengths are quantified by the parameters $A$ and $B$ that refer to coupling strengths within and between populations,
respectively. The functions $f$ and $h$ are $2\pi$--periodic that describe the coupling.  $N$ refers to the size of the populations. A schematic representation of this system is given in Fig. \ref{sys_schm}.

\begin{figure}
\begin{center}
\includegraphics[width=4.5cm]{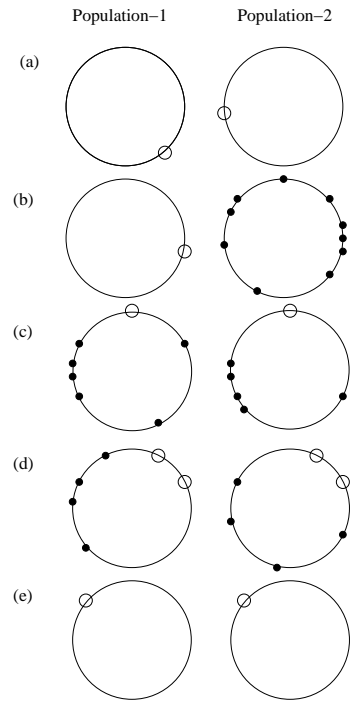}
\caption{Schematic representation of phase portraits of the states of synchronization in system (\ref{chim_del01}). (a) Individual synchronization in both the populations, (b) chimera, (c) GCC, (d) multi-clustered GCC, and (e) global synchronization. Open circles represent synchronized group of oscillators and the closed circles represent the desynchronized oscillators.}
\label{schm}
\end{center}
\end{figure}

Synchronization within a population can be characterized by using the complex mean field parameters $X^{(1,2)}+iY^{(1,2)}=r^{(1,2)}e^{i\psi^{(1,2)}}= \frac{1}{N}\sum_{j=1}^Ne^{i\theta_j^{(1,2)}}$. Here $r=(1/N)\sqrt{(\sum_j\cos\theta_j)^2+(\sum_j\sin\theta_j)^2}$ is also called the coherence parameter which measures the synchronization within a population. When $r=1$, there is complete synchronization in the system since in this state the phases of all the oscillators are the same. When $r$ takes values in between $0$ and $1$ there is either desynchronization or clustering in the population. In general, when $r$ takes a constant value, the corresponding state is a stable state, and when $r$ varies with time, the state is either a breather or an unstable state. However, synchronization between populations and global clustering cannot be characterized using these mean field parameters, since they represent average phases of the oscillators within a population. $\tau_1$ and $\tau_2$ quantify coupling delay within and between populations, respectively. A typical example of such a system is  the two groups of interacting neurons in the brain such as those in the cortex (say population 1) and the thalamus (say population 2) \cite{Jane:08b}. Another example of such a system is two layers (or columns) of interacting spin torque nano-oscillators, that need to be synchronized in order to generate coherent microwave power \cite{Mohanty:05}.

The occurrence of various synchronization states in system (\ref{chim_del01}) is schematically represented in Fig. \ref{schm}. Panel (a) represents a state of individual synchronization in the two populations where $r^{(1)}=1$ and $r^{(2)}=1$. However in this state the entrainment phases are different. Panel (b) is a chimera state where $r^{(1)}=1$ and $r^{(2)}<1$. Panels (c) and (d) represent the GCC and the multi-clustered GCC, respectively, where $r^{(1)}<1$ and $r^{(2)}<1$ for both the cases. Panel (e) represents a global synchronization state where $r^{(1)}=1$ and $r^{(2)}=1$, but the entrainment phases are equal unlike the case of individual synchronization (panel (a)).

\begin{figure}
\begin{center}
\includegraphics[width=8.5cm]{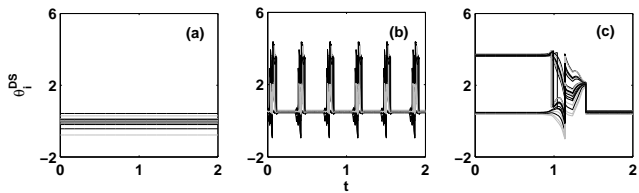}
\caption{Transition from (a) stable GCC through (b) periodic breather to (c) unstable GCC upon increasing the value of $\tau$ as (a) $\tau=0.85$, (b) $\tau=1.01$ and (c) $\tau=1.25$ for $\{f,h\}=\{\sin(\theta),\sin(\theta)\}$. Here $n=0$, $A=0.7$ and $B=0.4$. Time evolution of $\theta_i^{\mbox{\tiny{DS}}}$ are plotted. Black and grey lines represent oscillators in the first and second populations respectively. In this and the following numerical figures, we show only the desynchronized oscillators and the synchronized ones are not shown.}
\label{combined}
\end{center}
\end{figure}

We numerically simulated Eq. (\ref{chim_del01}) and discovered a motivating phenomenon of the existence of GCC states (note that system (\ref{chim_del01}) consists of two populations of \emph{identical} oscillators) as shown in Fig. \ref{Motiv}, and reported briefly in \cite{Jane:09b}. Interestingly enough we found that the coupling delay induces such phenomena where the system of identical delay--coupled populations splits into desynchronized and synchronized groups. This splitting can occur either within the populations or between the populations, depending upon the value of time delay for a given set of control parameters. The former represents the chimera and the latter is the GCC, as noted earlier.  Further, both the chimera and the GCC states need not be stable but they can either breathe or can be unstable as will be discussed later in Sec. \ref{Breat}.
\subsection{Numerical considerations}
\label{Numcon}
For all the numerical simulations we use Runge--Kutta fourth order routine with a time
step of 0.01 and we have also confirmed that the results are not affected by decreasing the time step below 0.01.
All the numerical figures depicted in this article are plotted after allowing a transient time of at least 2000 units, to reduce the likelihood of the presence of transients that may be mistaken for a dynamical behaviour. Actually, we have eliminated the first 2000 time steps before seeing the results and then the numerical plots are shown for small windows (100 or 200 time steps) towards the end of a simulation that lasted for 30000 time units.
In addition, we fix $N=32$ for numerical illustrations, although we have verified that the results are independent of the size of the system (for some details for $N=64$, see Sec. \ref{Breat}, Fig. \ref{NCheck}, and the corresponding discussion).
We also fix $\tau_1=n\tau_2=n\tau$, where $0\leq n\leq 1$. This condition pertains to the logic that the coupling delay within a population is always less than the coupling delay between the populations.

Since we also found that the chimera and GCC states need not be stable but can breathe depending upon the value of the coupling delay (details are given in the following sections),
we need to characterize such breathers. The mean field (coherence parameter) $r$ quantifies synchronization within a population and therefore it can be used
to quantify a breathing or unstable chimera. On the other hand, as mentioned earlier, global clustering/synchronization cannot be quantified using this order parameter.  Hence, in order to quantify a breathing GCC numerically, after allowing the transients, we separate out the synchronzied group from the desynchronized one. Note that, both the synchronized and desynchronized groups have oscillators from both the populations. 

We set a condition that $\theta_i|_{t=mT}=\theta_j|_{t=mT}$, for all ${i,j}$ (here $T$ denotes a particular time and $m=0,1,2,\ldots$ denotes discrete time steps of the numerical integration). The numerical equivalence for the above condition is upto 6 decimals in our calculation. Those oscillators which satisfy this condition are synchronized and remain synchronized for all times and are neglected so that we end up with the desynchronized group (that comprises oscillators from both the populations). While calculating the modified order parameter, we have to specify the minimum size of the synchronized group, that is we can choose how reasonably big a synchronized group can be. For example, if there are only two oscillators that have equal phases for all times, they cannot be considered as a group (given the large $N$) and hence we have to specify a minimum number of oscillators that satisfy the condition in order to be called as a synchronized group; the rest are considered as desynchronized oscillators. In our calculations for $N=32$ and $N=64$, we have taken this minimal value as $5$.

Let the size of the desynchronized group be $N^{\mbox{\tiny{DS}}}$. Now we can calculate the order parameter of this group as
\begin{eqnarray}
\label{dsr}
r^{\mbox{\tiny{DS}}}e^{i\psi^{\mbox{\tiny{DS}}}}=\frac{1}{N^{\mbox{\tiny{DS}}}}\sum_{j=1}^{N^{\mbox{\tiny{DS}}}}
e^{i\theta_j^{\mbox{\tiny{DS}}}},
\end{eqnarray}
where $N^{\mbox{\tiny{DS}}}=2N-N^{\mbox{\tiny{S}}}$. This order parameter $r^{\mbox{\tiny{DS}}}$ can be used to quantify both the chimera and GCC states and is also valid for cases where there exists more
than one synchronized cluster. Such multi-cluster states also occur for model (\ref{chim_del01}) which is discussed in Sec. \ref{Multi}.  Thus we define a cluster by imposing the condition mentioned in the previous paragraph, and identify the number of oscillators with the same phase, say $\theta_{i,j}=\Phi_j$, $i=1,2, \ldots m$, where $m$ is the size of cluster $j$. This process can be repeated for any number of clusters and each cluster can be characterized by the order parameter $r_n$ and the mean phase $\psi_n$. While specifying the size of clusters in multi-clustered GCC states, they have to be chosen to be relatively lower than the size of a synchronized group in a GCC state (where there is only one cluster).

\begin{figure}
\begin{center}
\includegraphics[width=6.2cm]{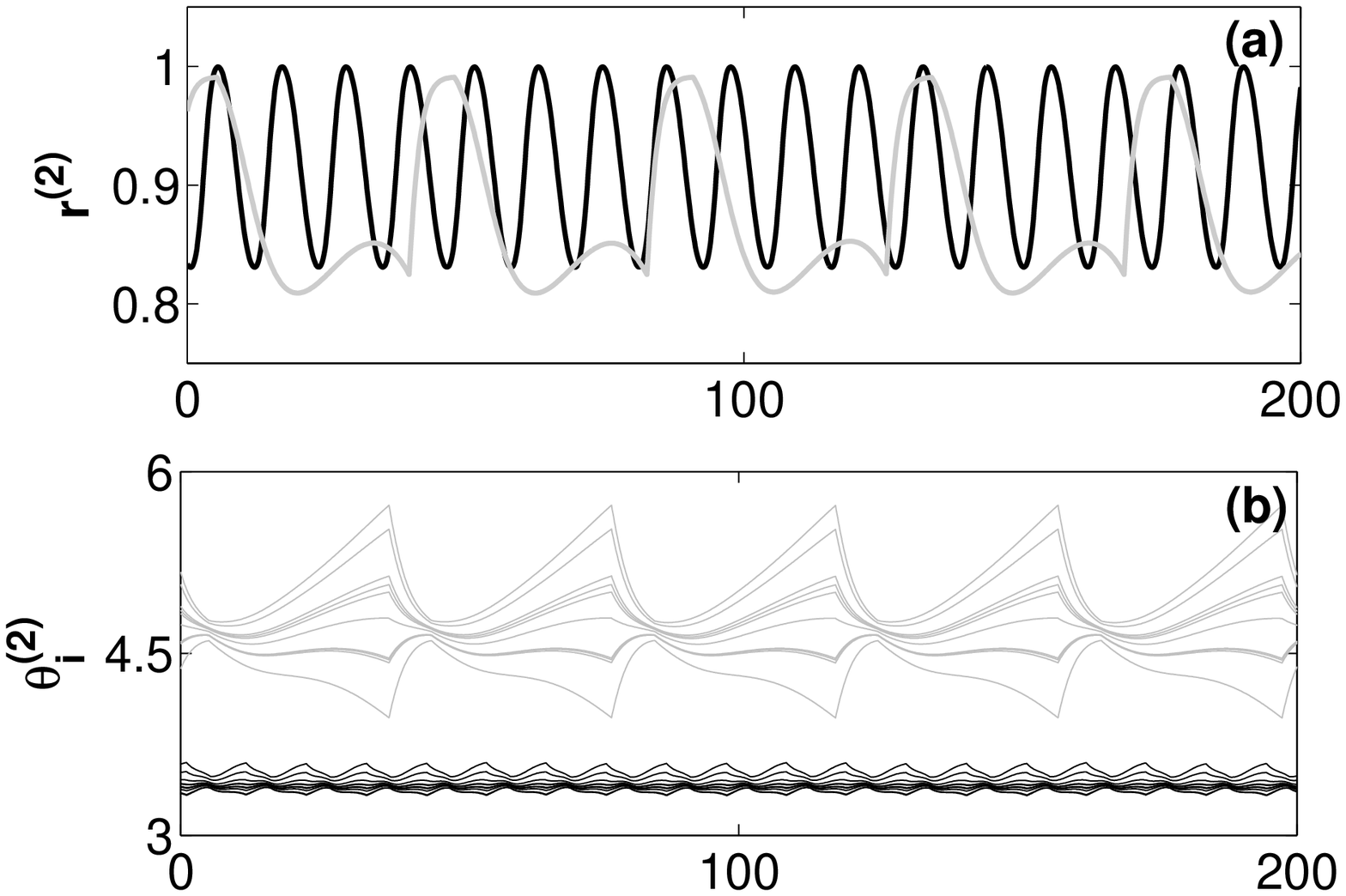}
\includegraphics[width=6cm]{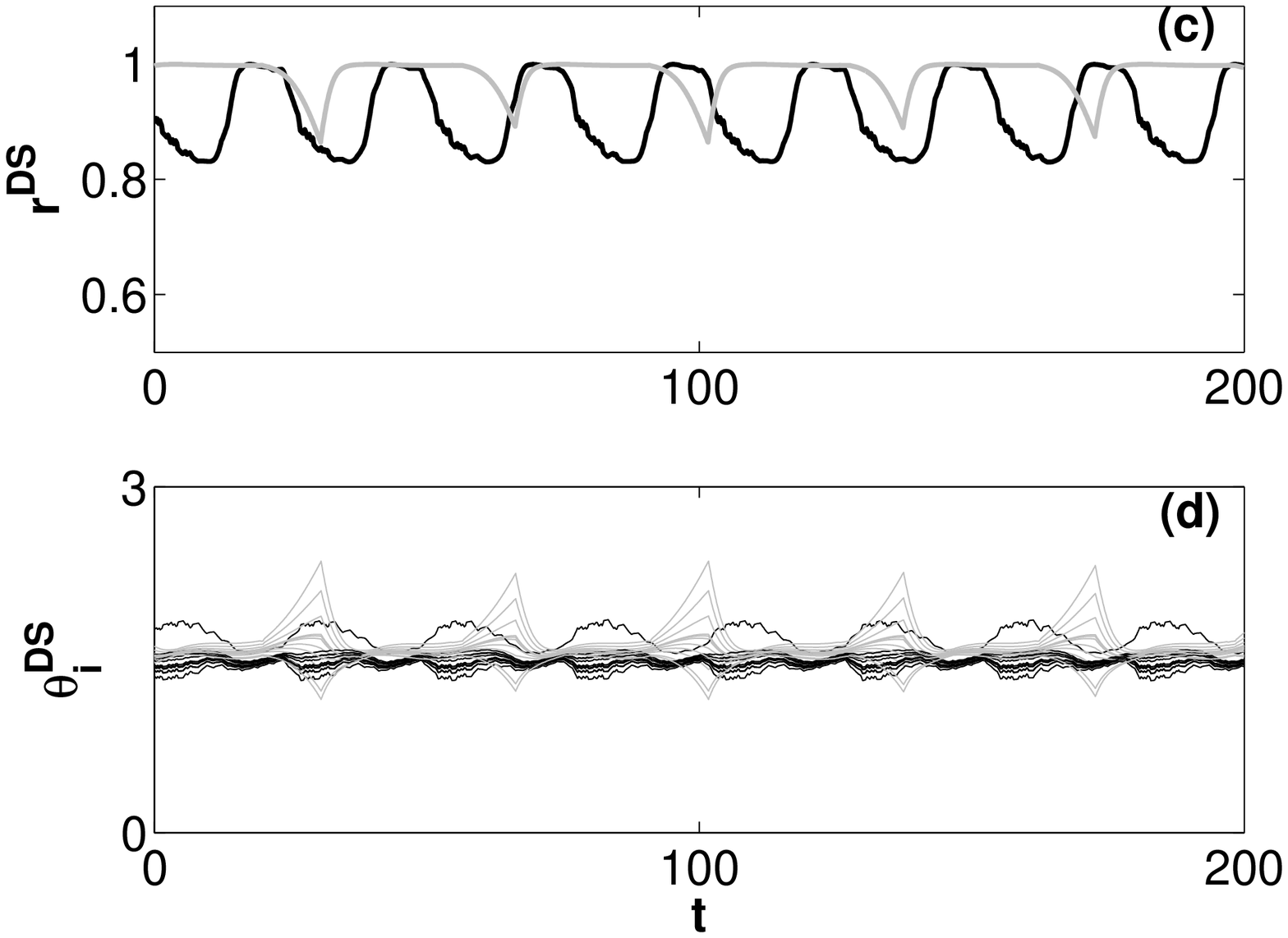}
\caption{Illustration of typical long (grey) and short (black) periodic breathers of the chimera ((a) and (b)) and GCC ((c) and (d)) states for $A=0.3$, $B=0.2$, $n=1$, $\{f,h\}=\{\sin(\theta),\sin(\theta)\}$. Here for chimera $\tau=2.3$ (long period) and $\tau=2.9$ (short period). For the GCC, $\tau=3.6$ (long period) and $\tau=4$ (short period). The
order parameters $r^{(2)}$ and $r^{\mbox{\tiny{DS}}}$, and the corresponding phases $\theta_i^{(2)}$ and $\theta_i^{\mbox{\tiny{DS}}}$ are plotted against time for chimera and GCC, respectively.}
\label{Rplot}
\end{center}
\end{figure}

\section{Breather and unstable states}
\label{Breat}
While a chimera or GCC is in the breather state, the phase of the synchronized group remains unaffected, but those of the desynchronized group fluctuate. The order parameter of the desynchronized group also fluctuates accordingly. The time delay parameter $\tau$ affects the stability of the chimera and the GCC states. Typical illustration of the occurrence of stable, breathing and unstable GCCs are shown in Fig. \ref{combined} where the time evolution of the phases of oscillators in the desynchronized group are plotted. For $n=0$, $A=0.7$, $B=0.4$ and $\{f,h\}=\{\sin(\theta),\sin(\theta)\}$, when $\tau=0.85$ (Fig. \ref{combined} (a)), the GCC state is stable. The desynchronized group of oscillators remain desynchronized, asymtotically. On increasing $\tau$ to $1.01$ (Fig. \ref{combined} (b)), we find that the GCC state loses its stability and ends up in what is called a breathing GCC state. In this state, the phase of the oscillators switches between synchronized (frequency suppressed) and desynchronized states. This breather illustrated in Fig. \ref{combined} (b) is a periodic breather as the switching process occurs periodically. Upon increasing $\tau$ further to $1.25$ (Fig. \ref{combined} (c)), one can visualize the example of an unstable GCC, where a desynchronized state loses its stability and a synchronized state becomes stable.

\begin{figure}
\begin{center}
\includegraphics[width=8cm]{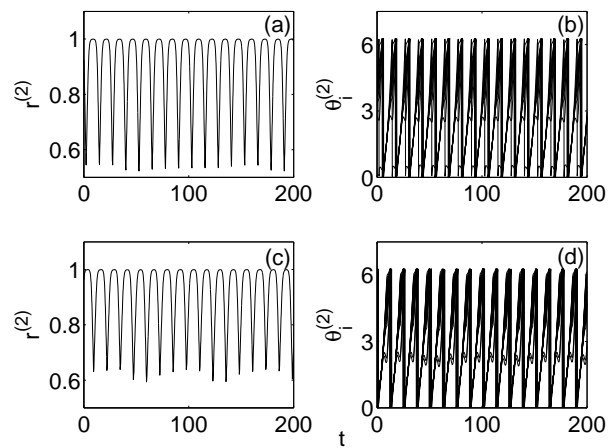}
\caption{Illustration of breathing chimera for two different values of the system size $N$ to demonstrate that the results are unchanged with $N$. Here  $\{f,h\}=\{\sin(\theta),\cos(\theta)\}$, $A=0.5$, $B=0.3$, $n=1$ and $\tau=1.9$. $N=32$ for (a) and (b) $N=64$ for (c) and (d).}
\label{NCheck}
\end{center}
\end{figure}

\begin{figure}
\begin{center}
\includegraphics[width=8cm]{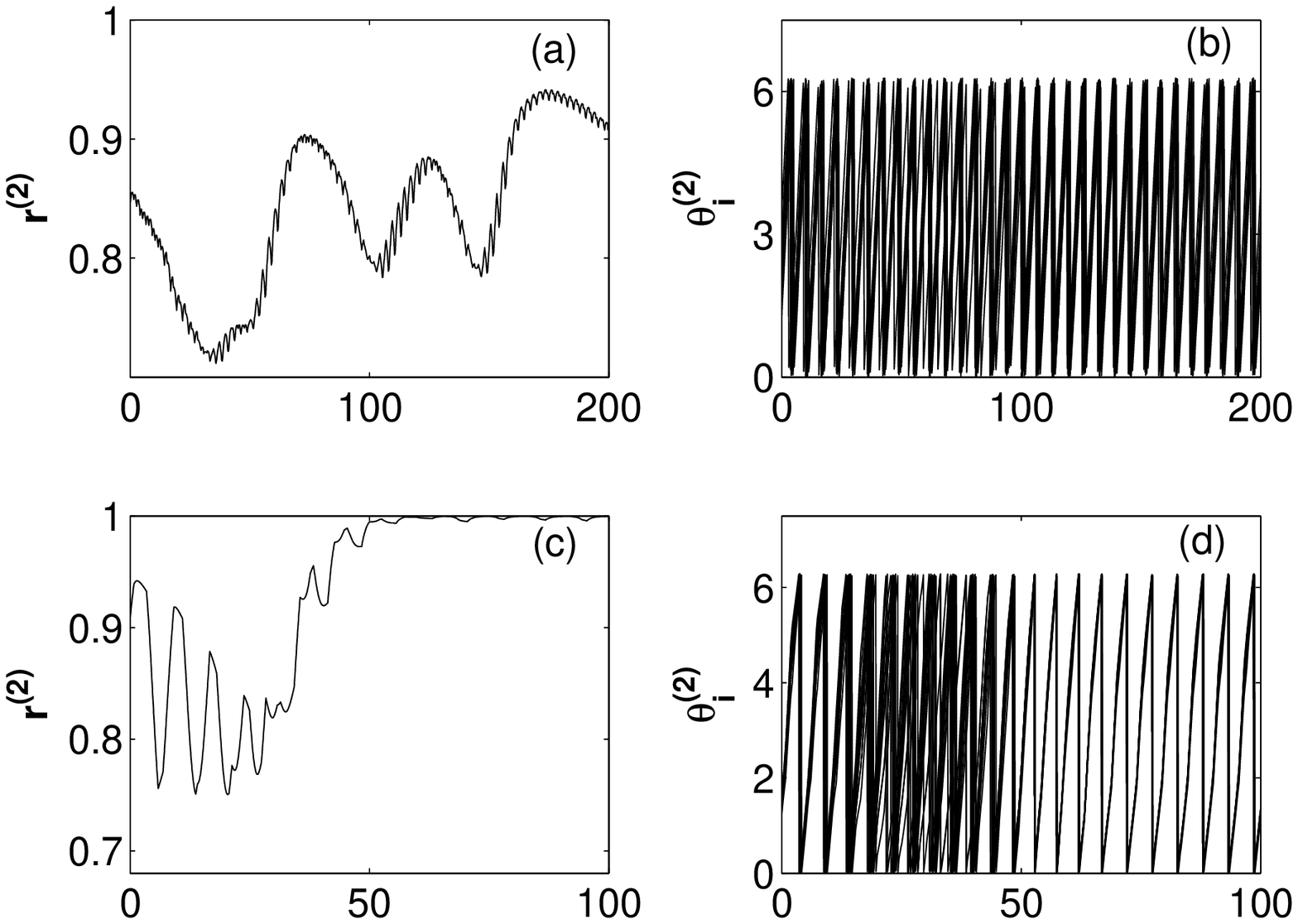}
\includegraphics[width=8cm]{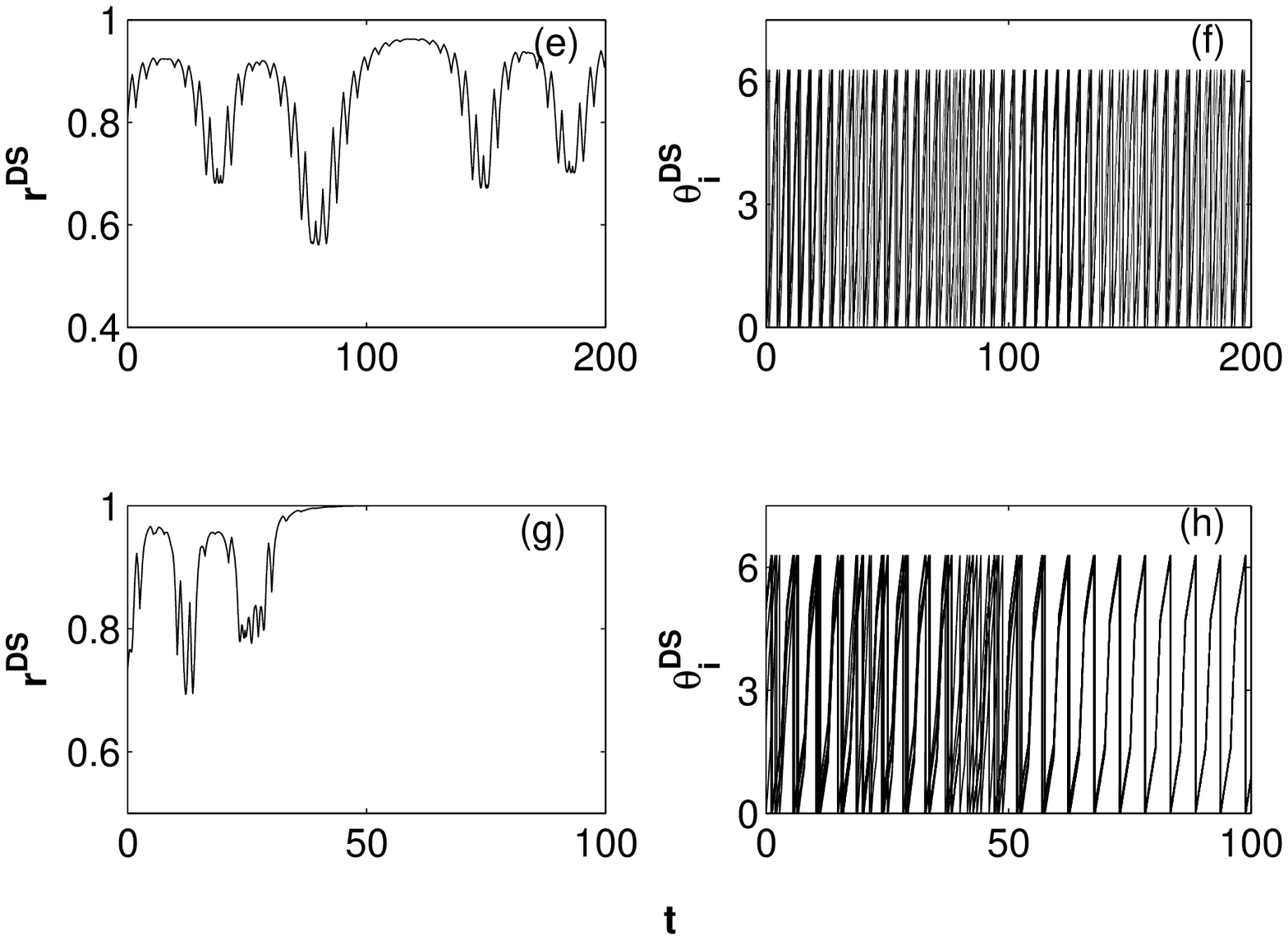}
\caption{Illustration of aperiodic and unstable chimera ((a)--(d)) and GCC ((e)--(h)) breathers for $\{f,h\}=\{\sin(\theta),\cos(\theta)\}$, $A=0.6$, $B=0.3$, $n=1$. Here $\tau=1.8$ for (a) and (b) (aperiodic chimera), $\tau=2.3$ for (c) and (d) (unstable chimera), $\tau=5$ for (e) and (f) (aperiodic GCC) and $\tau=4$ for (g) and (h) (unstable GCC).}
\label{LC1}
\end{center}
\end{figure}

There are different types of breathers in chimera and GCC states, namely periodic, aperiodic and unstable breathers (see Figs. \ref{Rplot} and \ref{LC1}). Typical short and long periodic breathers, defined in a relative sense,  are illustrated in Fig. \ref{Rplot}, for both the chimera and the GCC state. For the chimera, we plot the time evolution of the order parameter $r^{(2)}$ and the corresponding phases $\theta_i^{(2)}$ of the second population (since the first population is synchronized) in Fig. \ref{Rplot} (a) and (b), for two different values of $\tau$. For the GCC the order parameters $r^{\mbox{\tiny{DS}}}$ and the corresponding phases $\theta_i^{\mbox{\tiny{DS}}}$ are plotted for two different values of $\tau$ in Fig. \ref{Rplot} (c) and (d). For given values of system parameters, when $\tau=2.3$ we have a long period chimera breather. Upon increasing $\tau$ to $2.9$ we have a short period chimera breather. Long and short period GCC breathers occur on further increasing $\tau$ to $3.6$ and $4$, respectively. For the specific cases of the breathers illustrated in Fig. \ref{Rplot}, the desynchronized group switches between the states of frequency suppressed synchronization and desynchronization, corresponding to $r^{(2),\mbox{\tiny{DS}}}=1$ and $r^{(2),\mbox{\tiny{DS}}}<1$, respectively. In Fig. \ref{NCheck}, we have illustrated that the results are unchanged with the size of the system by plotting periodically breathing chimeras for two different values of $N$. In Fig. \ref{NCheck} (a) and (b) we have plotted the time evolution of the order parameter $r$ and the time evolution of the phases of the second population, respectively, for $N=32$ and in Fig. \ref{NCheck} (c) and (d) we have plotted the same for $N=64$.

Breathers need not be periodic; they can also be aperiodic. See Fig. \ref{LC1} for the illustration of a aperiodically breathing chimera and GCC states, where switching between different desynchronized states (the corresponding $r^{(2)}$ or $r^{\mbox{\tiny{DS}}}$ oscillates between $0$ and $<1$) occurs in an aperiodic manner. Plotted in (a) and (c) are the order parameters of the desynchronized population ($r^{(2)}$) for the chimera and in (b) and (d) are the corresponding time evolution of the phases. Similarly, in (e) and (g) the order parameters of the desynchronized group ($r^{\mbox{\tiny{DS}}}$) (obtained by neglecting the synchronized group, as explained earlier) are plotted with the corresponding time evolution of the phases in (f) and (h) that show aperiodic desynchronization windows.
The chimera and GCC states also become unstable depending on the value of the time delay parameter, where the oscillators in the desynchronized population/group remain desynchronized for a while, after which this state loses its stability and all the oscillators lock to one phase, that is, the desynchronized population/group becomes synchronized among themselves. The unstable chimera is shown in Figs. \ref{LC1} (c) and (d) and the unstable GCC is shown in (g) and (h) (after allowing 2000 time units for the transients to pass, as mentioned in Sec. \ref{Numcon}). As a consequence of the chimera or GCC state losing its stability, a two clustered synchronized state becomes stable. That is, when the chimera state becomes unstable, the individual synchronization state (as shown in Fig. \ref{schm} (a)) becomes stable, while for the unstable GCC, a state comprising of two groups of oscillators that are locked at two different phases becomes stable. The breather and unstable chimera/GCC states are not transient effects. As we have already mentioned in Sec. \ref{Numcon}, we have eliminated 2000 time steps before seeing the results and the numerical plots are shown for 100 or 200 time step windows towards the end of a simulation that lasted for 30000 time units. We also waited further more to see if aperiodic breathers collapse, but we find them to be steady dynamical states.

\begin{figure}
\begin{center}
\includegraphics[width=8cm]{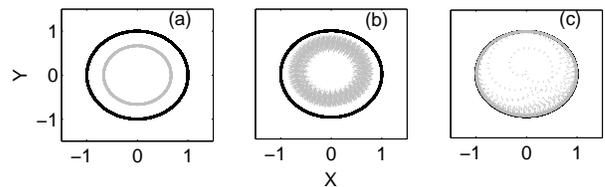}
\caption{Phase portraits showing the limit cycle of the synchronized population/group (the black line) and
desynchronized population/group (grey region). (a) stable chimera/GCC state (b) breathing chimera/GCC state and
(c) unstable chimera/GCC. Here $X=r\cos\psi$ and $Y=r\sin\psi$, where $r$ and $\psi$ are the mean-field parameters.}
\label{XY_LC2}
\end{center}
\end{figure}

The representation of stable, breathing and unstable chimera/GCC in the phase plane is shown in Fig. \ref{XY_LC2}.
The black line in Fig. \ref{XY_LC2} is the stable limit cycle attractor of the
synchronized population/group. This is always the same unit circle irrespective of the value of the entrainment
frequency of the synchronized population/group. The grey region represents the desynchronized population/group which
is stable in (a), breathing in (b) and becomes unstable in (c) of Fig. \ref{XY_LC2}.

\section{Multi-cluster states}
\label{Multi}

\begin{figure}
\begin{center}
\includegraphics[width=8cm]{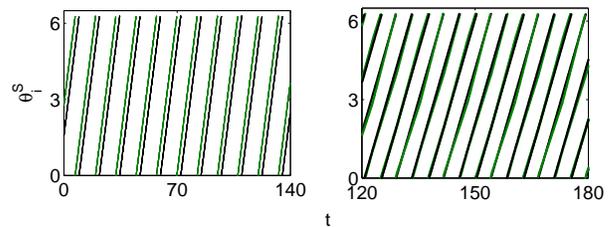}
\caption{(Color online) Occurrence of multi-cluster GCC states in system (\ref{chim_del01}). (left) Two-clustered ($\tau=2.8$) and (right) three-clustered ($\tau=3.7$) GCC states. Other parameter values are the same as those of Fig. \ref{Motiv}. Black and green (grey) lines represent oscillators in the first and second populations respectively. Here we have plotted only the synchronized group of oscillators.}
\label{Multi1}
\end{center}
\end{figure}

Multi-clustering is a phenomenon that commonly occurs while studying synchronization dynamics. For the
global clustering phenomenon studied in this paper, multi-clustered GCC states occur due to time delay.
Typical examples of two-clustered and three-clustered GCC states are illustrated in Fig. \ref{Multi1} where
the time evolution of the phases of the synchronized oscillators are plotted. While chimera states are a link
between sychronized and desychronized states \cite{Omel:08} (see also analytical explanation), the multi-cluster
GCC states are a link between the two (since we have two populations) types of chimera states. As we increase the
delay parameter $\tau$, the multi-cluster GCC state of the type shown in Fig. \ref{Multi1} and the GCC state occur for certain
values of $\tau$ in between the occurrence of chimeras. This is easily evident from Figs. \ref{Motiv} and \ref{Multi1}.
For the same values of all the other parameters as in Fig. \ref{Motiv}, these two figures show the occurrence of multi-clustered states in
between the chimera or GCC states for increasing values of $\tau$. When $\tau=2$, we have the chimera (with
population-1 synchronized and population-2 desynchronized, see Fig. \ref{Motiv} (left panel)). On increasing $\tau$ to $2.8$
we get a two-clustered GCC state as shown in Fig. \ref{Multi1} (left). One cluster contains oscillators only from the first population
and the other contains oscillators from the second
population. The desynchronized group of this state contains oscillators from both the populations. For $\tau=3.2$
we again get a chimera (with population-1 desynchronized and population-2 synchronized, not shown here) and for
$\tau=3.7$ we get a three-clustered GCC state shown in Fig. \ref{Multi1} (right). 

The above three-clustered GCC state
has three clusters each of which has oscillators from both the populations, which is different from the two-clustered
state. This difference in the two- and the three-clustered GCC states are due to the following reason: The two-clustered
GCC state occurs in between two chimera states and the three-clustered GCC state occurs in between a chimera and a GCC state.
Further, in between two chimera states and in between a chimera and a GCC state there can be more number of multi-clustered states,
depending upon the size of the cluster we choose. On increasing $\tau$ to $4$, we get
a GCC state as shown in Fig. \ref{Motiv} (right panel).
%\begin{widetext}
\begin{table*}
\caption{Occurrence of chimera and GCC for various values of the delay parameter $\tau$ (other parameter values are $\{f,h\}=\{\sin(\theta),\cos(\theta)\}$, $n=0$,
$A=0.4$, $B=0.6$)}
\begin{tabular}{|p{.7cm}|p{.8cm}|p{3.4cm}|p{11.2cm}|}
\hline
S.No.&Value of $\tau$ & State & Description \\
\hline
1&2.0 &  Chimera  &  Population-1 synchronized and population-2 desynchronized \\
\hline
2&2.8 &  2-clustered GCC & Two synchronized groups and one desynchronized group all containing oscillators from both the populations.  \\
 \hline
3&3.2 & Chimera &  Population-1 desynchronized and population -2 synchronized  \\
 \hline
4&3.7 &  3-clustered GCC & Three synchronized groups and one desynchronized group all containing oscillators from both the populations.  \\
 \hline
5&4.0 &  GCC &  One synchronized group and one desynchronized group all containing oscillators from both the populations.\\
 \hline
6&4.1 & Breathing GCC & One synchronized group and one desynchronized group that oscillates between different states.\\
 \hline
7&4.12 & Unstable GCC & One synchronized group and one desynchronized group that oscillates and becomes synchronized after a while.\\
 \hline
8&4.13 & Global Synchronization & One synchronized group.\\
 \hline
 \end{tabular}
\label{tab1}
\end{table*}
%\end{widetext}

Therefore we find that, on increasing/decreasing $\tau$ (starting with the state where both the populations are synchronized separately), the chimera first occurs, and further increase in $\tau$ causes switching between the two chimera states. Here the two chimera states necessarily mean, state-1 where population-1 is synchronized and population-2 is desynchronized and vice versa for state-2. This switching incorporates an intermediate chimera-like multi-cluster state where each of the synchronized clusters contains oscillators solely from one of the populations (as shown in Fig. \ref{Multi1} left panel). Further increase in $\tau$ results in multi-clustered GCC states, as shown in Fig. \ref{Multi1} (right), leading to stable GCC states. On increasing/decreasing $\tau$ further, this stable GCC state follows the
following sequence to end up in global synchronization: stable GCC, long--period breather, short--period breather, aperiodic breather
and unstable GCC leading to global synchronization. Further increase in $\tau$ from the global synchronization state
leads to a stable GCC by following the above mentioned route in the reverse order. These results are summarized in Table \ref{tab1} for a specific
set of parameters. As may be noted from the table, depending upon the values of the parameters $A$, $B$ and $\tau$, the behaviour repeats itself periodically. Therefore in order to visualize the occurrence of this series of phenomena one can either increase or decrease $\tau$ depending upon where we stand in the parameter space.

\section{Stability Of The Synchronized And Desynchronized States}
\label{Anal}

It is generally difficult to exactly pinpoint the occurrence of a GCC state in parameter space analytically. Our understanding so far \cite{Omel:08} reveals that chimera states are a natural link between synchronized and desynchronized states. In addition, our numerical evidence confirms that, on increasing $\tau$ for a given set of system parameters, chimera and GCC states occur periodically between stages of synchronization and desynchronization. Therefore one naturally needs to identify the boundaries that separate regions of synchronization and desynchronization and expect chimera/GCC states to occur near those boundaries.

In order to find the boundaries, we analyze system (\ref{chim_del01}) in the continuum limit $N\rightarrow \infty$. In this limit, a probability density for the oscillator phases can be defined as $\rho^{(1,2)}(\theta,t) d\theta$, which describes the number of oscillators with phases within $[\theta,\theta+d\theta]$ at time $t$. This distribution $\rho^{(1,2)}(\theta,t)$ obeys the evolution equation
\begin{eqnarray}
\frac{\partial\rho^{(1,2)}}{\partial
t}=-\frac{\partial}{\partial\theta}(\rho^{(1,2)}v^{(1,2)}),
\end{eqnarray}
where $v^{(1,2)}$ are given by
\begin{eqnarray}
\small
v^{(1,2)}&=&\omega - A\int_0^{2\pi}
%\nonumber \\ & \times &
f(\theta-\phi)\rho^{(1,2)}(\phi,t-\tau_1)d\phi \nonumber \\
&-&B
 \int_0^{2\pi}
 h(\theta-\phi)\rho^{(2,1)}(\phi,t-\tau_2)d\phi.
\end{eqnarray}
The functions $\rho^{(1,2)}(\theta,t)$ and $\{f,h\}$ are real and $2\pi$
periodic in $\theta$, so they can be expressed as Fourier series
in $\theta$, that is,
\begin{eqnarray}
\rho(\theta,t)&=&\sum_{k=-\infty}^{\infty}\rho(t)_ke^{ik\theta}, \quad  f(\theta)=\sum_{k=-\infty}^{\infty}f_ke^{ik\theta}\nonumber\\
h(\theta)&=&\sum_{k=-\infty}^{\infty}h_ke^{ik\theta}.
\end{eqnarray}
Substituting $\rho^{(1,2)}(\theta,t)$ and $\{f,h\}$ into the evolution
equation, we get
\begin{eqnarray}
\label{any05a}
\dot{\rho_l}^{(1,2)}&+&il\hat{\omega}\rho_l^{(1,2)}
\nonumber\\
&=&2il\pi\sum_{k=1}^{\infty}(a_k\rho_{l-k}^{(1,2)} +a_k^{\ast}\rho_{l+k}^{(1,2)}),
\end{eqnarray}
where
\begin{subequations}
\begin{eqnarray}
\label{7a}
\rho_{-l}^{(1,2)}=\rho_l^{\ast(1,2)}, \quad \hat{\omega}=\omega-(Af_0+Bh_0)
\end{eqnarray} and
\begin{eqnarray}
\label{7b}
a_k=(Af_k
\rho_k^{(1,2)}(t-\tau_1) + Bh_k \rho_k^{(2,1)}(t-\tau_2)).
\end{eqnarray}
\end{subequations}

Now, the linearized form of Eq.
(\ref{any05a}) reads as
\begin{eqnarray}
\label{any05}
 \dot{\rho_k}^{(1,2)}&=&-ik\hat{\omega}\rho_k^{(1,2)}+ika_k,
\end{eqnarray}
\noindent where the Fourier components for $|l|>k$ are neglected since
$l=\pm k$ are the only possible nontrivial unstable modes, $\rho_0=1/2\pi$
is the trivial solution corresponding to incoherence (desynchronization) and $f_k$ and $h_k$
are coefficients of the Fourier series of the functions $f$ and $h$. Now by considering only the non-trivial $k$th
Fourier mode $\rho_k$, and considering the linear stability of the desynchronized state $\rho_k=0$, we arrive at the eigenvalue
equation of that mode,
\begin{eqnarray}
\label{any05c}
(\lambda_{k}-\bar{A}e^{\lambda_k\tau_1}+i\omega_0)^2-\bar{B}^2e^{2\lambda_k\tau_2}=0,
\end{eqnarray}
where $\bar{A}=ikf_kA$,
$\bar{B}=ikh_kB$ and $\omega_0=k\hat{\omega}$. Equation (\ref{any05c}) leads to the pair of eigenvalue equations
\begin{eqnarray}
\label{any053c}
\lambda_k=\bar{A}e^{-\lambda_k\tau_1} \pm \bar{B} e^{-\lambda_k\tau_2} -i\omega_0.
\end{eqnarray}
These eigenvalues characterize the stability of the desynchronized state. This desynchronized state loses stability when the real part of the eigenvalue crosses the imaginary axis. Therefore, in order to obtain the stability boundary we assume $\lambda_k=-i\beta/\tau$, where $\beta$ is an arbitrary parameter, so that we can find the $k$th stability region in a parametric form as
%\begin{widetext}
\begin{eqnarray}
\label{chim_del02}
B=\pm kA\frac{|f_k|\cos(n\beta-\alpha_f)}{|h_k|\cos(\beta-\alpha_h)}; \; \;
\tau=\beta/[k(\omega_0 \nonumber \\ +A|f_k|\sin(n\beta-\alpha_f) \pm B|h_k|\sin(\beta-\alpha_h)]^{-1},
\end{eqnarray}
%\end{widetext}
where $f_k=-i|f_k|e^{i\alpha_{f}}$, $h_k=-i|h_k|e^{i\alpha_{h}}$  and $\tau_1=n\tau_2=n\tau$.
The overall stability of the desynchronized state is determined by the overlap of these domains for all the modes.

\begin{figure*}
\begin{center}
\includegraphics[width=8cm]{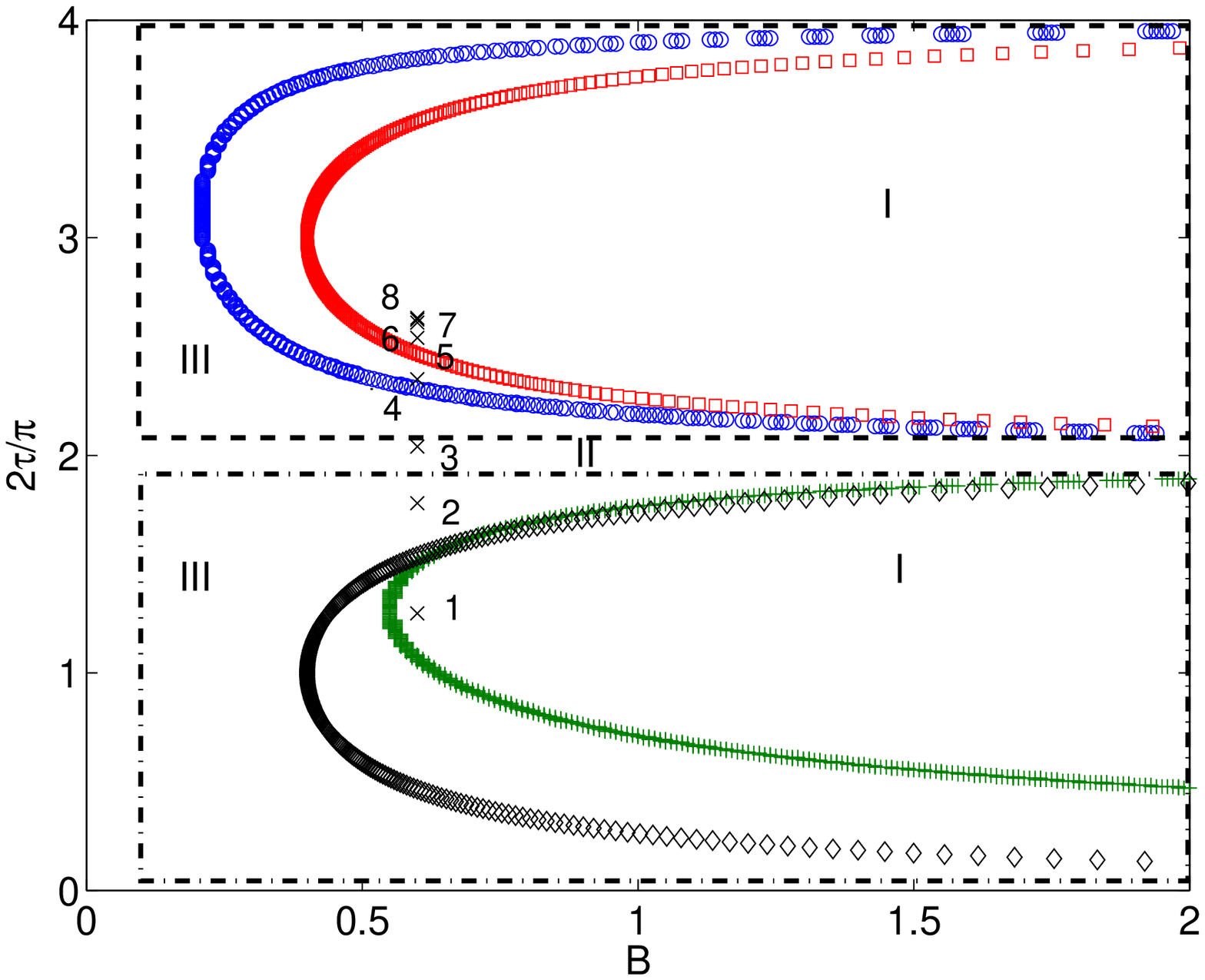}
\includegraphics[width=8cm]{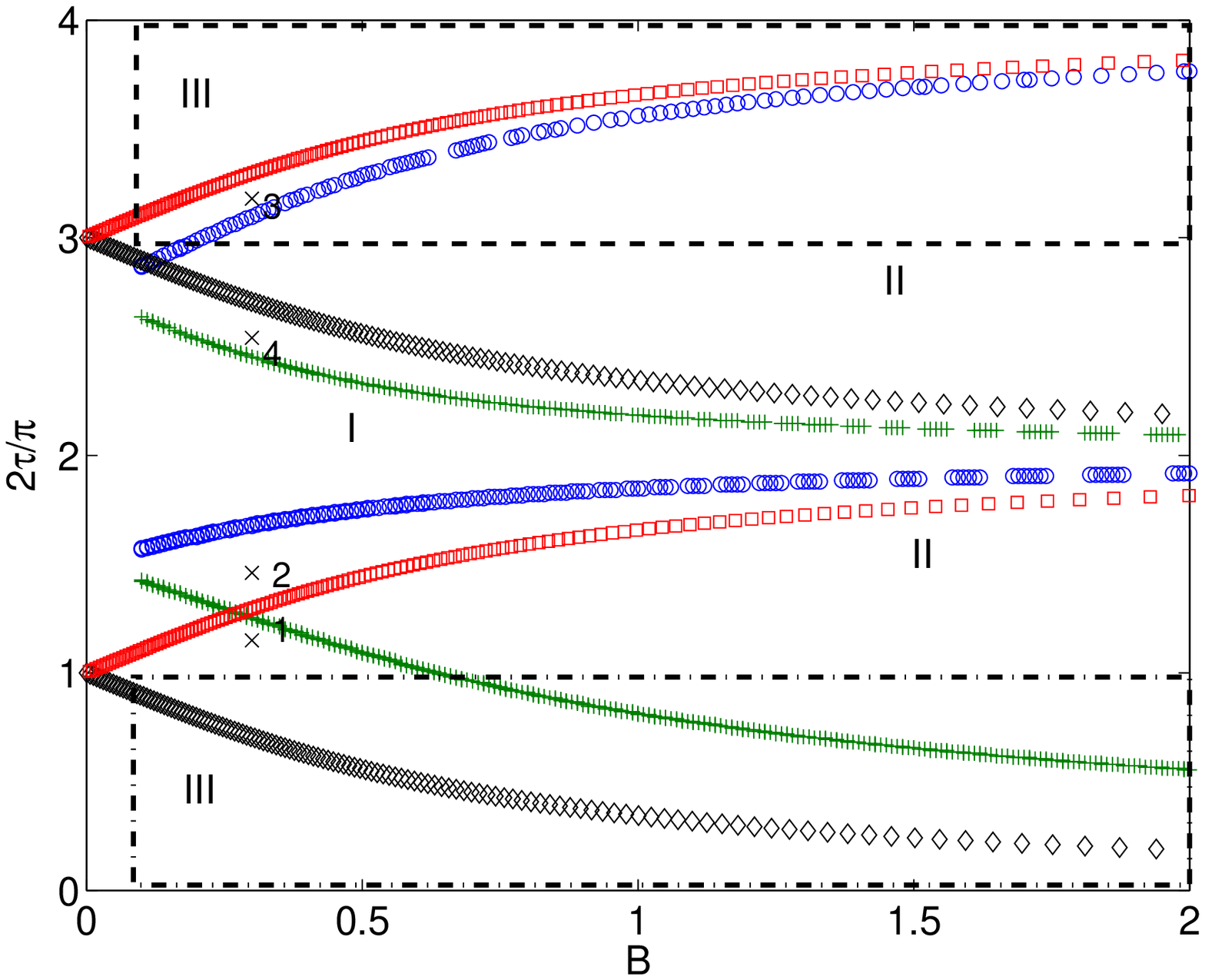}
\caption{(Color online) Theoretically obtained stability regions for $n=0$ (left) and $n=1$ (right),  I. Desynchronization, II. Synchronization of the populations individually and III. Global synchronization. Here $\{f,h\}=\{\sin(\theta),\cos(\theta)\}$. Boundaries with $\diamond$ and $\Box$ represent, respectively, $+$ and $-$ signs in equation (\ref{chim_del02}). These boundaries are the same as those for the in-phase and anti-phase synchronization states obtained from Eq. (\ref{chim_del04}). Dot-dashed and dashed curves correspond to in-phase and anti-phase synchronization states of (\ref{chim_del05a}). The regions bounded by $+$ ($+\hspace{-0.4cm}+\hspace{-0.4cm}+$) and $\circ$ ($\circ\hspace{-0.1cm}\circ\hspace{-0.1cm}\circ$) correspond to in-phase and anti-phase synchronization states of (\ref{chim_del05}). The markings with $\times$ denote the numerical examples; (left) 1-8 correspond to the states denoted in Table \ref{tab1} and (right) 1-4 correspond to panels (a), (c), (e) and (g) of Fig. \ref{LC1}. Note the numerical occurrence of the different chimera/GCC states near the analytical boundaries of synchronized and desynchronized states.}
\label{Btau1}
\end{center}
\end{figure*}

Now it is also of importance to investigate the stability of the synchronized state for
which we consider the solution $\theta_i^{(1,2)}=\Omega t$. With this solution,
system (\ref{chim_del01}) becomes
\begin{eqnarray}
\Omega=\omega-Af(n\Omega \tau)- Bh(\Omega \tau).
\end{eqnarray}
Along with this relation, the condition $Af'(n\Omega \tau)+ Bh'(\Omega \tau)>0$ should also be satisfied
in order that the synchronized state is stable. This provides the stability regime
\begin{eqnarray}
\label{chim_del04}
B=\frac{- Af'(n\beta)}{h'(\beta)}; \; \; \tau=\frac{\beta }{\omega-Af(n\beta)-Bh(\beta)},
\end{eqnarray}
where $\beta=\Omega \tau$. The parametric forms (\ref{chim_del02}) and (\ref{chim_del04}) separate the regions of
desynchronization and synchronization. For $\{f,h\}=\{\sin(\theta),\cos(\theta)\}$, these boundaries are plotted in Fig. \ref{Btau1}.

A homogeneous perturbation $\theta_i^{(1,2)}=
\Omega t+\Delta\theta^{(1,2)}$ pertaining to the case when all the phases remain equal while their rotation becomes nonuniform in time to the synchronization regimes leads to the following equations for stability
\begin{eqnarray}
\Delta\dot{\theta}_1&=&-(Af'(n\beta)+ Bh'(\beta))\Delta\theta_1 \nonumber \\ &&+Af'(n\beta)\Delta\theta_{1n\tau}- Bh'(\beta)\Delta\theta_{1\tau},\label{sta_con01}\\
\Delta\dot{\theta}_2&=&-(Af'(n\beta)+ Bh'(\beta))\Delta\theta_2 \nonumber \\
&&+Af'(n\beta)\Delta\theta_{2n\tau}+ Bh'(\beta)\Delta\theta_{2\tau},\label{sta_con02}
\end{eqnarray}
where $\Delta{\theta}_{(1,2)}=\Delta{\theta}^{(1)}\mp\Delta{\theta}^{(2)}$. The stabilty of the fixed point $\Delta{\theta}_{(1,2)}=0$ represents the global synchronization and individual synchronization of the populations. This is because, when $\Delta{\theta}_{(1)}=0$, $\Delta\theta^{(1)}=\Delta\theta^{(2)}$ and therefore both the populations are synchronized with the same entrainment phase. When $\Delta{\theta}_{(2)}=0$, $\Delta\theta^{(1)}=-\Delta\theta^{(2)}$ and hence both the populations are synchronized with different entrainment phase, the difference in the entrainment phase being $2\Delta\theta^{(1)}$. The stability conditions for (\ref{sta_con01}) for the cases $n=0$ and $n=1$ are \cite{DVS:07}
\begin{eqnarray}
\label{chim_del05a}
&&Bh'(\beta)>0, \;\;\;\;\qquad\qquad \qquad n=0,\nonumber\\
&&Af'(\beta)+ Bh'(\beta)>|Af'(\beta)- Bh'(\beta)|, \;\; n=1.
\end{eqnarray}
In both the above mentioned cases, the stability conditions for (\ref{sta_con02}) are
\begin{eqnarray}
\label{chim_del05}
&& \tau Bh'(\beta)+1>0, \;\; \;\qquad \qquad n=0,\nonumber\\
&&\tau(Af'(\beta)+ Bh'(\beta))+1>0, \;\; n=1.
\end{eqnarray}
For $\{f,h\}=\{\sin(\theta),\cos(\theta)\}$, the boundaries (\ref{chim_del05a}) and (\ref{chim_del05}) are plotted in Fig. \ref{Btau1}. The regions bounded by dot-dashed and dotted curves correspond to in-phase ($\theta_i^{(1)}-\theta_i^{(2)}=0$) and anti-phase ($\theta_i^{(1)}-\theta_i^{(2)}=\pm \pi$) synchronization states of (\ref{chim_del05a}), respectively. Similarly, the regions bounded by $\times$ and $\cdot$ correspond to in-phase and anti-phase synchronization states of (\ref{chim_del05}).

The stability boundaries between regions of the synchronized (both global and individual) and desynchronized states can be obtained using equations  (\ref{chim_del02}), (\ref{chim_del04}), (\ref{chim_del05a}) and (\ref{chim_del05}).  From these equations it becomes evident that the stability of the synchronized and the desynchronized states switch periodically between stable and unstable states on increasing/decreasing $\tau$, since $h$ and $f$ are $2\pi$ periodic. This also depends on the signs of $A$ and $B$. From Fig. \ref{Btau1}, it is obvious that on increasing $\tau$, the regions of synchronization and desynchronization alternate each other.  The chimera/GCC states can be expected near the stability boundaries of the synchronized and desynchronized states and hence the chimera/GCC states also repeat periodically on increasing $\tau$. This is evident from Fig. \ref{Btau1} where the numerical occurrence of the different chimera/GCC states, given in Table \ref{tab1} (markings 1-- 8 in Fig. \ref{Btau1} (left)) and Fig. \ref{LC1} (markings 1 -- 4 in Fig. \ref{Btau1} (right)) are near the analytical synchronization/desychronization boundaries. Note the (bistable) coexistence of globally synchronized state and desynchronized state (that is marking 8 in Fig. \ref{Btau1} (left) occurs inside Region I) due to the effect of time delay. Thus the stability analysis, while clearly pointing out the boundaries between synchronized and desynchronized states, also indicates the occurrence of chimera and GCC states.

\section{Applications}
\label{App}

It is well known that synchronization is not always desirable. For example, in the brain synchronization is desirable when it supports cognition via temporal
coding of information \cite{Singer:99,Jane:08b} while at the same time, it is undesirable when synchronization of a mass of neuronal oscillators occurs at a
particular frequency band resulting in pathologies like trauma, Parkinson's tremor and so on. Other examples include lasers and Josephson junction arrays \cite{Trees:05}, emission of microwave frequencies by coupled spin torque nano-oscillators \cite{Mohanty:05} where synchronization is desirable, while in the case of epileptic seizures \cite{Timmermann:03}, Parkinson's tremor \cite{Percha:05}, event related desynchronization \cite{Pfurtschelle:99}, or pedestrians on the Millennium Bridge \cite{Strogatz:01}, it is undesirable.

In the field of neuroscience, event-related synchronization and desynchronization of brain waves play a vital role in controlling higher level information processing, large scale integration and motor control and can be explained by models of diffusively coupled oscillators \cite{Jane:09c}. A thalamocortical model of three populations of neurons \cite{Jane:08b} to simulate the state of emergence from deep to light an{\ae}sthesia explains that successful coding of information and consciousness (on emergence from deep an{\ae}sthesia) is achieved by the occurrence of global synchronization between the thalamus and the cortex. In the corresponding experiment \cite{Bojan:07}, during deep an{\ae}sthesia $\delta$ waves (frequency in the range $1-4$Hz) occur with a high amplitude in the electroencephalogram (EEG). 

On emergence from deep to light an{\ae}sthesia the $\delta$ wave amplitude is dramatically decreased and $\theta$ (frequency in the range $4-8$Hz) waves begin to emerge in the EEG, however with a lower amplitude compared to that of the $\delta$ waves. The model demonstrates that, during deep an{\ae}sthesia, there is strong synchronization in the cortex giving rise to high amplitude $\delta$ waves. At the same time the neuronal oscillators in the thalamus are very poorly synchronized (which is similar to a chimera state) giving rise to very low amplitude $\theta$ waves and hence are not predominantly visible in the EEG. On emergence from deep to light an{\ae}sthesia some neuronal oscillators from the cortex desynchronize (and hence the dramatic decrease in the amplitude of $\delta$ waves) and combine with those in the thalamus to give rise to $\theta$ waves that are not as strongly synchronized as the $\delta$ waves during deep an{\ae}sthesia. This state may be considered as a state similar to a GCC state where neuronal oscillators from both the thalamus and cortex combine to form a synchronized group. The chimera and the GCC like states play a crucial role in the emergence from deep to light an{\ae}sthesia and also in blocking information transfer during deep an{\ae}sthesia (so that one does not feel pain) and for successful coding of information during light an{\ae}sthesia. Thus throughout the processes that take place in the brain the transmission delay in the propagation of neuronal signals causes the chimera and GCC like states to occur, that in turn facilitates the successful accomplishment of various tasks.

Another example prevails in the field of nanotechnology where the problem of synchronizing one or more populations of spin torque nano-oscillators at different columns to generate coherent microwave power is still open \cite{Mohanty:05}. By successfully modeling such a system using delay coupled populations of oscillators and exactly knowing where in parameter space the chimera/GCC occurs, one would be able to tune the system parameters, having the time delay parameter as the control parameter so to avoid the occurrence of a chimera/GCC. It will then be possible to stabilize the system in a state of complete synchronization.

\section{Summary}
\label{Sum}

To summarize, we have demonstrated that chimera and GCC states exist in delay coupled phase oscillator populations. A system of two identical, delay--coupled populations splits into two groups, one synchronized and the other desynchronized. The state is called chimera if one of the populations is synchronized while the other is desynchronized. On the other hand, the state is called a GCC if each group has a fraction of oscillators from both the populations. We have found that these states need not be stable always but can breathe periodically, aperiodically or become unstable, depending upon the value of coupling delay. In order to characterize the stable, unstable and breathing GCC states we have introduced a modified version of the order parameter that incorporates the mean of only the desynchronized group by neglecting the synchronized group of oscillators.

We also found that multi-clustered states exist as link between the chimera and GCC states, as the value of the time delay parameter is increased. We have provided analytical explanations of the observed effects on the basis of linear stability theory. Based on these results, we suggest that models that incorporate time delay serve as good candidates to explain many complicated natural phenomena as opposed to models without time delay. There are various methods to control synchronization (even its rate and velocity). One can choose regimes of synchronization or desynchronization depending upon coupling strengths, initial distribution of frequencies, the form of the coupling functions and so on. The message of the paper is that knowledge about the occurrence of chimera, GCC and multi--cluster states will help one to achieve a good control over synchronization and desynchronization in interacting populations of neurons, spin torque nano-oscillators and similar systems.

\section*{Acknowledgments}

The work is supported by the Department of Science and Technology (DST)--Ramanna program and DST--IRHPA research project, Government of India.

\end{document}